\documentclass[twocolumn,twocolappendix]{aastex701}

\usepackage{amssymb, amsfonts, amsmath,framed}
\usepackage{comment}
\usepackage{enumitem}

\usepackage{bm}

\usepackage[T1]{fontenc}
\usepackage{amssymb}
\usepackage{amsmath}
\usepackage{xspace}
\usepackage{graphbox}
\usepackage{graphicx,xparse,booktabs}
\newcommand{\Rcut}{$\mathcal{R}_{\mathrm{cut}}$}

\ExplSyntaxOn
\NewDocumentEnvironment{places}{mm}
 {
  \setlength{\tabcolsep}{0pt} 
  \dim_set:Nn \l_places_width_dim
   {
    (#1-\ht\strutbox-\dp\strutbox-0.2pt)/(#2)  
   }
  \begin{tabular}{r @{\hspace{2pt}} *{#2}{c}}  
 }
 {
  \end{tabular}
 }

\NewDocumentCommand{\place}{mm}
 {
  \seq_set_from_clist:Nn \l_places_images_in_seq { #2 }
  \seq_set_map:NNn \l_places_images_out_seq \l_places_images_in_seq { \places_set_image:n {##1} }
  \seq_put_left:Nn \l_places_images_out_seq
   {
   \scriptsize
    \begin{tabular}{c}\rotatebox[origin=c]{90}{\strut#1}\end{tabular}
   }
  \seq_use:Nn \l_places_images_out_seq { & } \\ \addlinespace
 }

\dim_new:N \l_places_width_dim
\seq_new:N \l_places_images_in_seq
\seq_new:N \l_places_images_out_seq

\cs_new_protected:Nn \places_set_image:n
 {
  \makebox[\l_places_width_dim]
   {
    \begin{tabular}{c}
    \includegraphics[
      width=\l_places_width_dim,
      height=\l_places_width_dim,
      keepaspectratio,
    ]{#1}
    \end{tabular}
   }
 }

\ExplSyntaxOff

\begin{document}

\title{Ultrahigh Energy Cosmic Ray Production in Binary Neutron Star Mergers}

\author[0000-0003-2417-5975]{Glennys R. Farrar}
\email{gf25@nyu.edu}
\affiliation{Center for Cosmology and Particle Physics, Department of Physics \\ New York University, New York, NY 10003, USA}

\begin{abstract}

Having previously argued that binary neutron star mergers are the principle source of ultrahigh energy cosmic rays~\citep{fBNS-prl25}, we exploit here the highly constrained initial conditions to make quantitative predictions for the cutoff energy of various nuclei.  UHECRs heavier than helium are accelerated in the magnetized turbulent outflow outside the jets to a rigidity $\mathcal{R}_{\rm cut} \equiv E_{\rm cut}/eZ \approx 6-9$ EV, consistent with the measured value $\mathcal{R}_{\rm cut} = 6.3^{+6.3}_{-2.3}\,$EV from fitting data.  This agreement strengthens the case that BNS mergers are the main site of UHECR production. 

The jets may accelerate protons and/or helium to cutoff energies $\approx 11.5$ and $\approx 35$ EeV, respectively.  Such a jet component and its spallation products could explain the indication of a secondary light population at higher energy found in the analysis of~\citet{muf19}.  The relative abundances of different elements and the total energy in UHECRs per merger event will become calculable, pending advances in our understanding of the mechanism of ion uptake into the acceleration process and input from nuclear physics experiments.  

This scenario implies that each neutrino above 1 PeV is co-directional with a gravitational wave arriving $\approx 1$ day earlier, and that the highest energy UHECRs have masses heavier than iron. 

\end{abstract}


\section{Introduction}
\label{sec:Intro}

In the six decades since the first solid evidence for UHECRs above 100 EeV$\,\equiv10^{20}\,$eV~\citep{Linsley}, many suggestions have been made for astrophysical systems capable of accelerating charged particles to such extreme energies.  Proposed sources include the jets of Active Galactic Nuclei (AGN) and of long gamma ray bursts (GRBs) from core-collapse of massive stars, jets or shocks of tidal disruption events (TDEs), shocks of giant radio lobes or extragalactic large scale structure, and a diversity of other systems;  see, e.g.,~\citet{Anchordoqui18} for an overview.

Initially it was widely assumed that the highest energy UHECRs must be protons both on account of that element's abundance but also because heavier nuclei could disintegrate in an intense radiation field.  However data now clearly shows that the composition of extragalactic cosmic rays evolves from light (protons and helium) to iron or heavier~\citep{augerDNNPRD25} as the energy increases from a few EeV to the highest observed energies exceeding 200 EeV.  Whether the highest energy UHECRs have masses beyond iron is presently not known; the composition has only been reported to $10^{19.75}$ eV due to the small number of events at higher energy, moreover there is an intrinsic uncertainty in mapping the air shower observables to the primary UHECR's mass due to inadequacies in particle physics models of the hadronic interactions that govern the air shower development~\citep{augerDNNPRD25,augerCompICRC25}.

In spite of the uncertainty as to the absolute mass scale, a robust and unexpected feature of the UHECR data has emerged: the {\it spread} of nuclear masses at any given energy is very narrow.  The data is consistent with the spectrum at production depending only on the rigidity, $\mathcal{R}\equiv E/Ze$ for relativistic particles.  This behavior implies the charge increases in proportion to the energy and is known as a Peters cycle.  It is not surprising since interactions with a magnetic field depend on a particle's rigidity, so the ability of a system to retain charged particles and continue to accelerate them places a limit on their rigidity.  However not only does the mass track the energy, as follows when the spectrum depends primarily on rigidity, but {\it the rigidity distribution itself is very narrow}.  This can be seen in Fig.~4 of~\citet{bf23} showing the rigidity distribution of Auger events with energy above 8 and above 32 EeV.  Besides the narrowness of the distributions for different nuclei, one sees that the mean rigidity, 3.5 EV and 4.5 EV is practically constant: a factor-4 change in energy threshold only changes the mean rigidity by $\approx 20$ percent. 

For UHECRs to have a narrow rigidity distribution requires two distinct, unforeseen properties to be true:  1) The individual UHECR accelerators must produce UHECRs with a narrow range of rigidities, i.e., they must have a hard spectrum and sharp cutoff, and 2) the ensemble of UHECR sources must have near-identical values of their rigidity cutoffs.

An early manifestation of the narrowness of the rigidity spectrum was the shockingly hard spectral indices found by Auger in their combined fits to the spectrum and composition.  Diffusive shock acceleration (DSA) predicts a spectrum of the form $\mathcal{R}^{-p} {\rm exp}(-\mathcal{R}/\mathcal{R}_{\rm max})$~\citep{ProtheroeStanev99}, with $p\approx 2$, but the Auger combined fits with this functional form have consistently returned unphysical power law indices, sometimes as extreme as $p = -1.8$ (see Table 1 of ~\citet{augerCombFitJCAP23}).\footnote{This is for a ``broken exponential" spectrum, which changes from pure power law to pure exponential at $\mathcal{R}=\mathcal{R}_{\rm cut}$. Using the DSA power times exponential form generally requires a harder index, as seen in Table 6 of~\citep{augerCombFitJCAP17}.  Note that the effective power-law index hardens by up to one unit when escape from the magnetized surroundings of the accelerator is taken into account, but remains in tension with data by about one unit~\citep{UFA15,mf23}. Also note that the spectrum quoted is always that escaping from the source since the observed spectrum is modified by energy losses and spallation during extragalactic propagation.}  

The tension between theoretically predicted and observed spectral indices is resolved using the spectrum predicted for magnetized turbulence, determined from PIC simulations by~\citet{cfm24} to be
\begin{equation*}
\phi(\mathcal{R}) \propto \mathcal{R}^{-p}\, {\rm sech}\left[ (\mathcal{R}/\mathcal{R}_{\rm cut})^2 \right]~,   
\end{equation*}
where $p\approx 2$ as for diffusive shock acceleration. Fitting the Auger data with the much sharper sech cutoff gives a better fit to the combined spectrum and composition data than the exponential cutoff~\citep{cfm24}; the derived best-fit spectral index is $p = 2.1$, consistent with theory.  In a nutshell, unphysically-hard spectral indices found in combined fits to spectral and composition data are an artifact of trying to fit a narrow rigidity distribution with a soft, exponential cutoff.  An important task for the particle-acceleration community is to use particle in cell (PIC) simulations to measure the shape of the cutoff for diffuse shock acceleration.  If the analytic prediction of~\citet{ProtheroeStanev99} is confirmed, the UHECR data will disfavor diffuse shock acceleration as the mechanism which accelerates UHECRs.

However it is not enough for individual sources to have a sharp cutoff and narrow rigidity distribution -- the observed narrow UHECR rigidity distribution requires individual UHECR sources to have the {\it same} cutoff, to sufficient accuracy to not spread the overall rigidity distribution too much.  The implications of this were first critically examined by~\citet{foteini+23}, who used the relation between the maximum rigidity to which a system can accelerate UHECRs ($\mathcal{R}_{\rm max} \propto B \times L$, where $B$ is the rms magnetic field and $L$ the size of the system), and its bolometric luminosity which is proportional to its Poynting flux $\propto (B \times L)^2$, to constrain the population-variance of the sources.  ~\citet{foteini+23} showed that even if individual sources produce a sharply peaked rigidity distribution, the observed luminosity functions of AGN, long GRBs, TDEs and other studied systems are too broad for consistency with the narrow UHECR rigidity distribution.  

The inability of AGN, long GRBs, TDEs and other previously suggested UHECR accelerators such as astrophysical shocks, to satisfy the ``standard source'' requirement, prompted the suggestion~\citep{fBNS-prl25} that binary neutron star mergers are the source of UHECRs.  Amongst candidate UHECR source classes, BNS mergers appear uniquely capable of accounting for the standard source phenomenon, because the post-merger magnetic field is generated by a gravitationally-driven dynamo~\citep{Kiuchi24} and the mass range of BNS systems is narrow; see~\citet{fBNS-prl25} for more detailed discussion.  Another attractive feature of the BNS-merger scenario for production of UHECRs is that it provides a natural explanation for the population of extremely high energy events significantly above 100 EeV~\citep{fBNS-prl25} -- namely that those UHECRs originated as r-process nuclei produced in the BNS merger, with a similar rigidity distribution as other UHECRs but a higher charge than Z=26.

The BNS-merger-origin of UHECRs was advanced in~\citet{fBNS-prl25} on empirical grounds.  In the present paper, we examine the mechanism of UHECR production and acceleration in a BNS merger.  ~\citet{Kiuchi24} showed through very high resolution GRMHD simulations including neutrinos, that the post-merger magnetic field is generated by a gravitational dynamo.  Thus differences from one BNS merger to another, e.g., in the initial magnetic fields of the neutron stars, are insignificant in determining the magnetic field of the post-merger system.  It also means that the ~\citet{Kiuchi24} simulation provides a robust initial condition for the magnetized turbulent outflow of {\it any} BNS merger, as we exploit here.  We follow the turbulent magnetized medium, initialized to the~\citet{Kiuchi24} simulation, as it expands and accelerates UHECRs.  We quantitatively predict the cutoff energies of different nuclei, and give an approximate analytic treatment of the spectral shape of UHECRs escaping the source environment.  We estimate the total energy of UHECRs created per unit volume per unit time and discuss the calculations necessary to predict the relative abundance of different UHECR nuclei.  The general arc of the presentation is to start with analytic analyses of simple robust features, then move to more qualitative discussions of other aspects.

Section~\ref{sec:locus} reviews the regions of the post-merger system where UHECR acceleration may occur, concluding that the magnetized turbulent ejecta at polar angles outside the jets is the principal and possibly the exclusive source of UHECRs.  Section~\ref{sec:composition} describes in qualitative terms how successive stages of evolution of the outflow play specific roles in producing the observed spectrum and composition.  In subsequent sections we show that some features can be quantitatively treated, some can be approximately estimated, and others require detailed simulations and theoretical developments beyond the scope of this work.

Section~\ref{sec:MagTurbAcc} discusses acceleration by magnetized turbulence and shows that the required conditions apply in the bulk outflow outside the jets.  Initially, synchrotron cooling limits the maximum energy attained by the ions in the outflow, but as the system expands the field strength drops as $r^{-3/2}$ until synchrotron losses no longer inhibit reaching the maximum rigidities the system can produce. In Sec.~\ref{sec:spectrum} we identify the locus in the outflow at which the highest rigidities are achieved and determine the value of the rigidity cutoff.  In the process, we uncover a potential valuable probe of this UHECR production scenario.  Section~\ref{sec:escape} gives an approximate analytic treatment of how escape from the ejecta further shapes the spectrum of cosmic rays.

In Sec.~\ref{sec:jet} we provide an estimate of the cutoff rigidity of protons and alpha particles accelerated in the jet, under the approximation that at the relevant large distance from the center, the jet outflow can be modeled as magnetized turbulence.  In Sec.~\ref{sec:energetics} we calculate the total mass of nuclei accelerated per BNS merger, show that the energy in UHECRs per merger times the merger rate is sufficient to explain the UHECR energy production rate within present uncertainties, and discuss steps to tighten this constraint.   
In Sec.~\ref{sec:probe} we discuss tests and probes of the BNS-merger scenario for UHECR origins, including determining the characteristic time delay of extremely high energy (EHE) neutrinos with energies above one PeV relative to the gravitational wave (GW) produced by the merger itself; observation of such coincidences would be a definitive proof of the BNS merger scenario and may be possible with next-generation GW and neutrino detectors.  
A summary and list of major outstanding challenges is given in Sec.~\ref{sec:summary}. 

\section{Jets, sheath or bulk outflow?}
\label{sec:locus}
Three regions in the post-merger system are candidates for being the locus of UHECR acceleration: the jets, the bulk outflow at larger angles, and the sheaths of the jets, i.e., the interface between jets and bulk outflow. Up to now, the most popular mechanism for UHECR acceleration has been diffusive shock acceleration in jets.  But on empirical grounds, as explained below, acceleration in the jets is at most a secondary contributor to UHECRs.  

At small polar angles, neutrino radiation suppresses the neutron abundance and the particles produced are protons and/or alpha particles.  The composition becomes heavier with increasing polar angle.  At large angles, nuclei with $Z$ and $N\geq 20$ are dominant; as explained by~\citet{Perego+LightElems22} "the production of elements between Li and K is negligible in all relevant kinds of ejecta. This very robust feature is due to the presence of $Z= N= 20$ magic nuclear numbers, which prevents seed nuclei formed in the iron group region to reach elements below calcium."  

The UHECR data is unambiguous that the composition of extragalactic CRs starts light (p and He) around the ankle at a few EeV and becomes progressively heavier with energy, reaching up to Fe~\citep{augerSDXmax17,augerDNNPRL25,augerDNNPRD25} and perhaps beyond.  Heavy nuclei can be spallated and photodisintegrated to produce p and He, but not vice-versa. This empirical evidence thus shows that the jets of BNS mergers are not responsible for the highest energy cosmic rays.  
We therefore focus particularly on UHECR acceleration in the magnetized turbulent outflow outside the jet.  We find (Sec.~\ref{sec:MagTurbAcc}) that the cutoff rigidity and the power law spectral index for nuclei accelerated in this region are in good agreement with observations.  
  
While the jets cannot be the main source of UHECRs, the jets may provide a distinct source of p and/or He.  The cutoff energies for p and He accelerated in the jets are estimated in Sec.~\ref{sec:jet}.  The results are consistent with jet acceleration producing the separate light component found to contribute of order 10\% of the energy above 10 EeV in the analysis of ~\citet{muf19}. This analysis needs to be repeated using the improved composition determination now possible. 

Another possible locus of acceleration is the mixing layer in the sheath of the jet, at an angle of a few to 10 degrees~\citep{Sironi+Shear21}. The recent study by~\citet{PaisPiran+24} of the late-time evolution of the jet in a BNS merger shows that such a mixing layer does form, but probably at too small angle to produce the needed heavier mass UHECRs based on the composition as a function of polar angle found by~\citep{Perego+LightElems22}.  Moreover, the total energy available for UHECR production in this mixing layer is most likely too small to account for the observed UHECR flux.  

Yet another possible acceleration site, potentially relevant for lower energy cosmic rays and neutrinos, is at the shocks created by fall-back onto the expanding ejecta, as explored in ~\citet{Decoene+19} and~\citet{Rossoni+24}, where further references to the literature can be found.

\section{Predicting the composition distribution}
\label{sec:composition}

For the reasons discussed above, we focus initially on UHECRs produced in the magnetized ejecta outside the jets.  Simulations of nucleosynthesis in the post-merger outflow of a BNS merger report a rather wide range of total mass in the ejecta, $(0.4-10)\times 10^{-3}\, M_\odot$ depending on the equation of state and whether a long-lived remnant precedes formation of a black hole~\citep{Curtis+Radice22,Perna+24}.  Fortunately this and most other uncertainties do not affect our analysis of UHECR acceleration, as the discussion below shows.   For a good general overview of BNS mergers and application to observations of GW170817 see~\citet{MetzgerThompsonQuataert18}.  

The radius of the outflow, $r$, increases linearly with time, with a velocity $v_{\rm ej}\approx 0.1-0.2\,$c.  The simulation of ~\citet{Rosswog+BNS14} followed the evolution for 100 yr and found that, to excellent approximation, the expansion is homologous.  We therefore adopt the homologous-expansion approximation, 
which determines how the field strength evolves starting from where it is initialized with the~\citet{Kiuchi24} simulation.  

Nuclei form in the expanding outflow when the temperature drops enough that photodisintegration becomes subdominant to nucleosynthesis; this occurs when the radius of the ejecta is $r\approx 10^{10}\,$cm, about 1$\,$s after collapse.  At large angles, outside the jet region, the synthesized nuclei are essentially all heavy, with $Z,N\geq 20$~\citep{Perego+LightElems22}.  As the expansion continues, a fraction of the nuclei become energized by acceleration in the magnetized turbulence, initially by reconnection~\citep{Comisso19,cfm24}.  The probability of uptake into the acceleration process 
(often called ``injection" in the literature)
 can depend on the charge and mass of the ions, as well as the temperature, density and other properties of the turbulence.  These dependencies, and also the overall efficiency of uptake, have not yet been mapped out; they need to be measured in PIC simulations.  Since this is not yet known, we are unable at present to predict the composition of the initial low energy CR nuclei and the absolute normalization of the UHECR spectrum. 

Once a nucleus reaches a moderate energy, collisions with other nuclei cause spallation and breakup.  
As seen from both~\citet{Canto+HeavyIon20} and~\citet{ZeitlinFrag+11}, collisions between heavy nuclei produce secondaries with a broad spectrum of mass ratios.\footnote{
Experimental data on the collision of two r-process nuclei is sparse, but we have direct experimental evidence of the broad range of fragment masses from~\citet{ZeitlinFrag+11} who used nuclear beams ranging from $^{14}$N to $^{45}$Mg with energy per nucleon of 290-1000 MeV on a range of targets including C, Al, Sn and Pb.  The review article by~\citet{Canto+HeavyIon20} discusses both experiments and modeling. Given the very high charges involved, behavior can be quite different from what is familiar from lower-mass nuclear collisions.  Based on Figs.~19 and 21 in~\citet{Canto+HeavyIon20}, one can estimate that breakup occurs for center-of-mass kinetic energies 20-100 MeV and above, and that the typical cross section for breakup is $\gtrsim 100\,$mb.}  
 Hence we can be confident that CRs with a wide distribution of masses is available for subsequent acceleration. 

The ultimate composition of UHECRs is governed by a complex, evolving reaction network; suppressing indices labeling the atomic numbers of the nuclei and the momentum distributions, 
the rate per nucleus of a breakup-inducing collision is a sum over $\Gamma_{\rm bu} = f_{\rm bu} \Gamma_{\rm col} $, where $f_{\rm bu}$ is the fraction of encounters which cause breakup, while $ \Gamma_{\rm col}(r) = n(r) \, \sigma \, v_{\rm th}(r) $ is the collision rate with $n,\,\sigma,\, v_{\rm th}$ being the number density, cross section and mean relative velocity respectively.  The density $\approx 1\,$s after collapse to a black hole, when the ejecta is sufficiently cool for nuclei to form, is comparable to the density of water;  thus the collision rate of nuclei having the required energies is large.  The breakup probability into different channels determines the composition distribution of CRs available for subsequent acceleration. Further study of this topic is left to the future.

\begin{figure}[t]
    \includegraphics[trim={1.3in 1.4in 1.2in 1.5in},clip,width=0.46\textwidth]{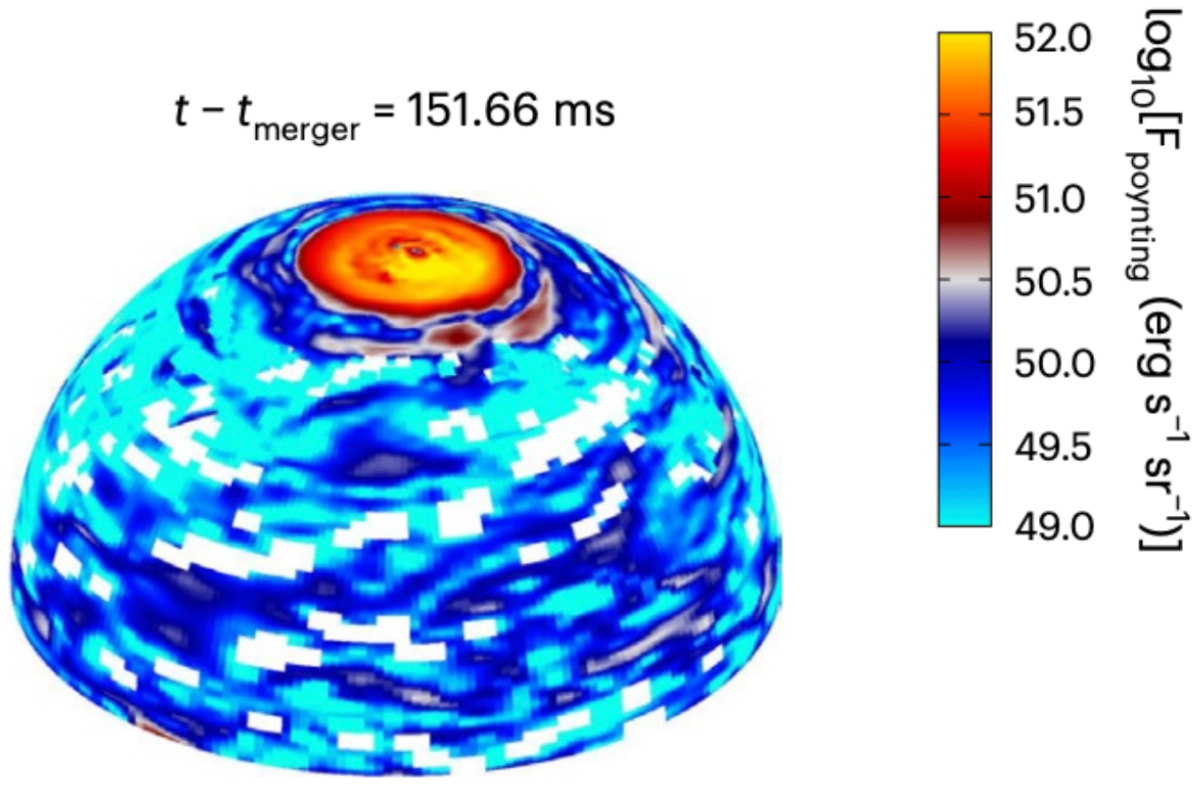}
   \caption{The Poynting flux on a sphere of $r= 500$ km, 150 ms post-merger, in the high-resolution neutrino-radiation-GR-MHD simulation of a BNS merger reported in Fig.~4 of~\cite{Kiuchi24}. } 
    \label{fig:KiuchiPoynting}
\end{figure}

\section{Acceleration in magnetized turbulence}
\label{sec:MagTurbAcc}   

The magnetized outflow outside the $\approx \!10^\circ$ jet cone in a BNS merger is an example of an environment with magnetized turbulence.  The term ``magnetized turbulence" refers to a region in which the energy density in a turbulent magnetic field is greater than the kinetic energy density in particle random motions, $\epsilon_{K}$. 

A valuable resource for our analysis is the very high resolution neutrino-GRMHD simulation of the merger of two equal mass (1.35 $M_\odot$) neutron stars by~\citet{Kiuchi24}.  These authors showed that the magnetic field produced by a BNS merger is generated by a {\it gravitationally-driven dynamo}, and they measured the properties of the field in different regions.  Figure~\ref{fig:KiuchiPoynting} reproduces Fig.~4 of ~\citet{Kiuchi24}, showing the Poynting flux on a sphere of radius 500 km, 151.66 ms after the merger.  Outside the jet, the Poynting flux is typically $\approx  10^{49.5} \,{\rm erg \, s^{-1}\,sr^{-1}}$. The characteristic field strength corresponding to this Poynting flux is $B\approx\! 3.3 \times 10^{12}$ G, so the mean energy density in the magnetic field $u_B \approx 4.2\times 10^{23}\,{\rm erg \,cm^{-3}}$ at the initial configuration of our analysis.  \citet{Kiuchi24} also find the spectrum of turbulence is consistent with Kolmogorov. Estimating the angular size of the coherence scale of the magnetic turbulence from Fig.~\ref{fig:KiuchiPoynting}, gives a coherence length of $l_{\rm coh} \approx r/3 $, 
the same relationship found by~\citet{cfm24} in their static box simulations.  With the assumption of homologous expansion discussed above, these initial conditions determine the magnetic field strength and coherence length during subsequent stages of the expansion.

The temperature of the ejecta in the rest frame of the outflow following collapse is found in simulations to be of order 10 GK~\citep{Curtis+Radice22}, corresponding to a typical particle kinetic energy $\approx 1$ MeV;  nuclei are non-relativistic in the outflow rest frame, while the bulk velocity of the ejecta is $\approx (0.1-0.2)c$~\citep{HamidaniKiuchiIoka20}.  The total ejecta mass is uncertain, but is generally found to be $\lesssim 10^{-2} M_\odot$. Thus $\epsilon_K \lesssim u_B$ at $r=500$ km.  
The magnetic energy density $u_B$ evolves $ \propto r^{-3}$, since the total energy in the magnetic field is approximately constant during expansion.  The kinetic energy per particle evolves during expansion as $p^2/(2M) \sim r^{-2}$, so the kinetic energy density decreases as $\sim r^{-5}$ and becomes even more subdominant to the magnetic energy density as the radius increases. Thus the outflow is appropriately described as magnetized turbulence.

It has recently been shown using state-of-the-art PIC simulations~\citep{cfm24}, that acceleration in magnetized Kolmogorov turbulence produces a sharply cutoff spectrum of UHECRs of the form 
$
\phi(E) \propto E^{-p}\, {\rm sech}\left[ (E/E_{\rm cut})^2 \right]$.   
The response of a relativistic charged particle to a magnetic field depends on its rigidity, $\mathcal{R} \equiv E/(Ze)$, so for greater generality we write
\begin{equation}
\label{eq:specR}
\phi(\mathcal{R}) \propto \mathcal{R}^{-p}\, {\rm sech}\left[ (\mathcal{R}/\mathcal{R}_{\rm cut})^2 \right]~.   
\end{equation}
Fitting the data in their PIC simulations, ~\citet{cfm24} found the dependence of $\mathcal{R}_{\rm cut}$ on the rms random field strength and its coherence length to be:
\begin{equation}
\label{eq:Rcut}
    \mathcal{R}_{\rm cut,\, EV} = (3 \times 10^{-16}) \,0.65\, B_{G} \, l_{\rm coh}
\end{equation}
where we have inserted the numerical prefactor to convert from erg to EV. 
In the setup of~\citet{cfm24}, where the system is a rectilinear static box, $p \approx 2.4$ within the magnetized turbulence.  The escape from the accelerator in that geometry was estimated to harden the spectrum of escaping CRs to $p \approx 2.1$.  In Sec.~\ref{sec:escape} below we will estimate how the spectrum of escaping CRs is shaped in the expanding geometry relevant to BNS mergers.  

 The Auger collaboration~\citep{augerCombFitJCAP23} fit their data with a variety of functional forms for the cutoff and a free power law index.  With the ${\rm sech}\left[ (\mathcal{R}/\mathcal{R}_{\rm cut})^2 \right]$ cutoff of Eq.~\eqref{eq:specR} they found ${\rm log}_{10} \mathcal{R}_{\rm cut,V}  = 18.78\pm 0.01$ (from Table 8; statistical errors only).  \citet{cfm24} also fit to published Auger data, adopting the generalized framework of~\citet{UFA15} to describe escape from the source, and found ${\rm log}_{10} \mathcal{R}_{\rm cut,V} =  18.8\pm0.02$ (statistical errors only) and power law index $p = 2.1$.   

\section{UHECR spectrum in the merger outflow}
\label{sec:spectrum}

\begin{figure}[t]
\includegraphics[width=0.48\textwidth, trim=0cm 0cm 0cm .45cm, clip=true]{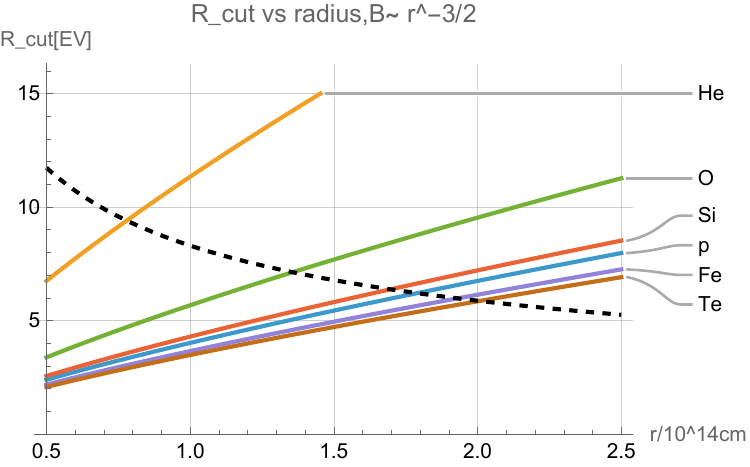}
    \caption{Black, dashed: rigidity cutoff $\mathcal{R}_{\rm cut}$ as a function of radius from Eq.\eqref{eq:Rcut}, with field initialized to the~\cite{Kiuchi24} simulation and evolving with expansion as $B\propto r^{-3/2}$ and $l_{\rm coh} = r/3$. Colored, solid: Maximum rigidity for various nuclei as a function of radius, consistent with $t_{\rm synch} = t_{\rm accel}$. The radii where these intersect is denoted $r_{\rm crit}$ for each composition.
    \label{fig:synchlim}}
\end{figure}

In the absence of synchrotron inhibition, the cutoff rigidity in magnetized turbulence would be given by Eq.~\eqref{eq:Rcut}, leading to $\mathcal{R}_{\rm cut}$ decreasing $ \sim r^{-1/2}$.  The value of $\mathcal{R}_{\rm cut}$ from  Eq.~\eqref{eq:Rcut},  initializing the magnetic field in the outflow by the~\citet{Kiuchi24} simulation, is shown by the dashed black line in Fig.~\ref{fig:synchlim}.\footnote{
The absolute normalization has an $\mathcal{O}(1)$ uncertainty since the numerical prefactor 0.65 appearing in Eq.~\eqref{eq:Rcut} was measured in a static system~\citep{cfm24} and could differ in detail in the spherically expanding merger outflow.} 
However, the maximum rigidity which can be achieved at a given radius for a given nuclear species is limited by the requirement that $t_{\rm synch} \gtrsim t_{\rm accel}$, as detailed in Appendix A.  The resultant $\mathcal{R}_{\rm max}(r)$ is shown for a representative set of nuclei, by the colored solid lines in Fig.~\ref{fig:synchlim}.  The intersection of the dashed and solid lines gives $\mathcal{R}_{\rm cut}$ for each species. Thus, $\mathcal{R}_{\rm cut } = 
6.2,\,9.4,\,7.1,\,6.4,\,6.0$ and 5.9
EV for p, He, O, Si, Fe and Te, with the radius of maximum rigidity, $r_{\rm crit}$ reached at 
(1.8, 0.8, 1.4, 1.7, 1.9, and 2.0)$\,10^{14}\,$cm, respectively, using the 0.65 prefactor in Eq.~\eqref{eq:Rcut}.  

This range of estimated \Rcut\ values is to be compared to the results of fits to Auger data quoted above: $\mathcal{R}_{\rm cut,EV} = 6.0\pm 0.1 \rm(stat.)\pm 0.8 \rm(syst.)\,$EV~\citep{augerCombFitJCAP23} and $\mathcal{R}_{\rm cut,EV} = 6.3^{+6.3}_{-2.3}\,$EV~\citep{cfm24};  the slight difference between these fits is due to the treatment of systematic uncertainties.  
Auger has an overall energy scale uncertainty of about 14\% and the absolute mass scale (ln$A$) has a comparable uncertainty from the hadronic interaction modeling.  
Both fits assume a single Peters Cycle with common rigidity cutoff, so are not precisely applicable to the prediction here, but the general agreement is encouraging.

Better understanding of the evolution of the composition out to $\approx \! 10^{14}\,$cm is needed (see Sec.~\ref{sec:composition}) to know whether lighter nuclei such as p, He and O are a significant component of the  outflow and accelerated directly, or are only formed later through spallation and breakup of heavier UHECRs as they interact with photon fields or matter on the rest of their path to Earth.  In the latter case, the produced nuclear fragments such as He and O will characteristically have a similar \Rcut\ to other nuclei in the approximation that Bethe-Heitler energy losses can be neglected.  However if 
He and O and other light nuclei are present in the ejecta in time to be accelerated directly, the analysis of this section shows they will be accelerated at smaller $r$, to up to 50\% higher values of \Rcut\ than protons and heavier nuclei.  

When PIC simulations have determined the appropriate prefactor in Eq.~\eqref{eq:Rcut} for the case of an expanding outflow, future ability to measure the absolute composition will enable not only the central value of \Rcut\ to be used to test the theory, but detailed differences between the spectra of different compositions will give a detailed test of this scenario and can in principle distinguish where the intermediate mass nuclei are accelerated.

\section{Spectrum escaping the acceleration region}
\label{sec:escape}

Equation~\eqref{eq:specR} gives the rigidity spectrum inside the expanding outflow.  The \emph{observed} spectrum consists of CRs which escape the outflow and are further processed by energy losses and transmutation during propagation to Earth.  
The treatment of energy losses and composition evolution during extragalactic propagation is straightforward using publicly available codes such as CRPropa and SimProp.  However accurately predicting the spectrum of cosmic rays escaping the expanding magnetized medium is highly non-trivial. In a complete treatment one must take into account that during the time it takes a CR to escape, the ejecta radius and coherence length increase, and the field strength decreases, with respect to their initial values.  While an analytic approximation for this can be developed, a proper treatment must also take into account the CR's diffusion in energy, the back reaction of the UHECRs on the magnetic field depleting its energy~\citep{cfm24}, as well as the UHECR's interactions with the ambient photon field while it escapes~\citep{UFA15}.  Incorporating these effects will not impact the key result of this paper, namely estimation of \Rcut, but are needed to predict the low energy spectral shape and the total energy in escaping UHECRs.  These topics are beyond the scope of the present study; a rough treatment is given in Appendix~\eqref{app:escape}.    

\section{p and He accelerated in the jet}
\label{sec:jet}
 
Let us now consider acceleration in the jet region, at small polar angles. Ideally, we would use PIC simulations of diffusive shock acceleration in the BNS merger jets and predict the form of the rigidity cutoff, but these simulations have not been done. However after the substantial expansion to radii where the field strength is small enough that synchrotron emission no longer limits CR acceleration, the initial jet may have evolved such that it is adequately described by magnetized turbulence rather than as a system of colliding internal shocks.  (For discussions of DSA and acceleration in magnetized turbulence see, e.g., ~ \citet{DruryDSA83} and~\citet{LemoineMagTurb25}, respectively).  

In this approximation, we can estimate the rigidity cutoff of protons and He accelerated in the evolved jet by the same method as in Sec.~\ref{sec:spectrum} for the large angle outflow.  The relative abundance of p and He in the jet is sensitive to the nuclear equation of state and can range from p dominance to He dominance for different allowed NSEoS (see Fig.~2 of~\citet{Perego+LightElems22}) so we consider both options below.  The normalization of jet-component UHECRs  relative to the outflow component depends on the contentious problem of baryon loading of jets and is beyond the scope of this work.

\begin{figure}[t]
\includegraphics[width=0.48\textwidth, trim=0cm 0cm 0cm 0cm, clip=true]{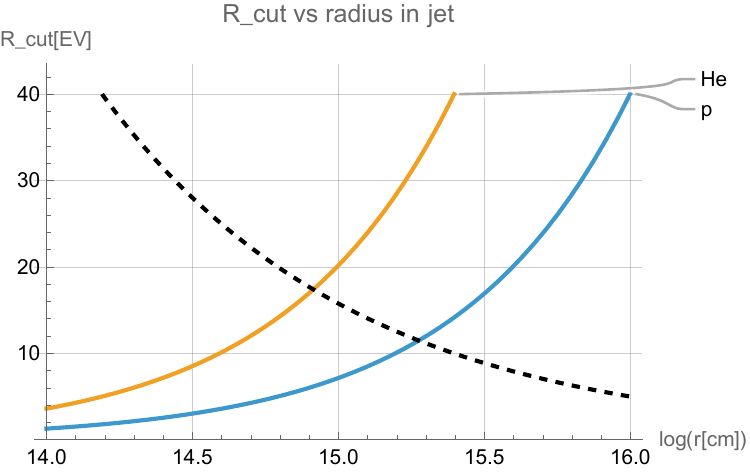}
    \caption{As in Fig.~\ref{fig:synchlim}, but initializing for the jet, where $B$ is 10 times larger and $l_{\rm coh} \approx 0.2\, r$.}   
    \label{fig:synchlimjet}
\end{figure}

Figure~\ref{fig:KiuchiPoynting} reproduced from~\citet{Kiuchi24} shows that the Poynting flux in the jets is about a factor 100 greater than in the  outflow at larger angles considered in the previous section, so the initial field strength is a factor 10 higher. The coherence length is approximately the jet radius, so given the 12$^\circ$ opening angle, we take $l_{\rm coh}= 0.2 \,r$.  With these parameters, the analysis detailed above balancing the timescales for synchrotron emission and acceleration, leads to Fig.~\ref{fig:synchlimjet} and the estimate that a  distinct light component of p and/or He would have $\mathcal{R}_{\rm cut} \approx 11.5$ (17.4) EV, respectively, reached at critical radii of $1.9\, (0.8)\times 10^{15}\,$cm.  Given the uncertainty in the estimate of $l_{\rm coh}$ at the relevant distance and the approximation of applying the magnetized turbulence analysis to the jet, these estimates have $\mathcal{O}(1)$ uncertainties.   A nucleon resulting from  photo-disintegration of jet-accelerated helium en route to Earth has $E_{\rm cut}\approx 9\,$ EeV, ignoring extragalactic losses, compared to 11.5 EeV for a proton directly accelerated in the jet or 6 EeV in the large angle outflow. 

Such a distinct higher energy light component of UHECRs may be present in the data.  ~\citet{muf19} report that their fit to UHECR data is improved by $\approx 5\sigma$ by including a secondary pure proton component with $E_{\rm max} \approx 10^{19-19.2} = 10-15$ EeV, carrying 3-10\% of the energy of CRs above 10 EeV,  with the range of values reflecting different hadronic interaction models; see Table II and the Appendix of~\citet{muf19}.  The latest analysis of composition evolution~\citep{augerDNNPRL25,augerDNNPRD25} has much smaller statistical uncertainties than available in 2019 thanks to development of the Deep Neural Network analysis enabling surface detector events to be used.  However until hadronic interaction models are able to correctly describe the muon content, shower development and zenith angle dependence of the observations, conclusions from such analyses must be treated as tentative.  Future analyses of UHECR data should allow for a higher energy, non-Peters cycle component consisting of protons and/or He, as well as allowing for a component heavier than Fe.  

If a distinct light component accelerated in the jets is found, its spectrum could provide a probe of whether the acceleration mechanism is diffusive shock acceleration (DSA) or acceleration in magnetized turbulence.  According to the analytic leaky-box analysis of~\citet{ProtheroeStanev99}, diffusive shock acceleration produces an exponential or softer spectral cutoff.  As shown in~\citet{cfm24}, the overall UHECR data prefers the sharper sech cutoff.  However a distinct subdominant light component could have an exponential cutoff~\citep{muf19}.  The DSA spectral cutoff should be measured in PIC simulations to determine if the~\citet{ProtheroeStanev99} result holds up in some or all geometries of the shock relative to the field. 

\section{Total UHECR mass \& energetics}
\label{sec:energetics}

From the results of previous sections we can estimate the total mass of nuclei accelerated to ultrahigh energy, to get an idea what fraction of the nuclei synthesized in the merger ultimately wind up as UHECRs. 
The total energy carried by the UHECRs is the sum over nuclear species of the energy carried by nuclei of each species. Integrating the approximate escaping spectrum shown in Fig.~\ref{fig:escSpec}, gives a mean rigidity of about 0.2 EV, for a mean energy per nucleon $\approx \!0.1\,$EeV. An upper limit on the total energy in UHECRs is the equipartition value, i.e., half the initial magnetic field energy, $\mathcal{O}(10^{51})\,$erg. At a typical energy per nucleon of 0.1 EeV,  the total mass of the nuclei accelerated to UHE is $\lesssim 5 \times 10^{-12}\, M_\odot$.  Given estimates of the total mass of the ejecta of $(0.4-10)\times 10^{-3}\, M_\odot$~\citep{Curtis+Radice22,Perna+24}, we arrive at an order-of-magnitude estimate that approximately one billionth of the nucleons present in the outflow are accelerated.  A complete theory of the evolution of the outflow and the mechanism of CR uptake needs to explain this value which, when better determined, will be a highly non-trivial test of our understanding. 

We next consider the overall energy budget.  As summarized in Sec.~\ref{sec:MagTurbAcc}, the very high-resolution neutrino-radiation-GRMHD simulation of the merger of two 1.35$M_\odot$ neutron stars by~\citet{Kiuchi24} allows us to initialize our treatment of the outflow 150 ms post-merger at a radius of 500 km, where the mean value of the Poynting flux outside the jet is 
$F_{\rm Poyn}\approx 10^{49.5} \,\rm{erg \,s^{-1} sr^{-1}}$.  
The jets occupy just a few percent of the solid angle, so the total energy in the magnetized outflow is $\approx 4 \pi F_{\rm Poyn}\, T$, where $T$ is the lifetime of the metastable system prior to its collapse to a black hole. Simulations for equal mass neutron stars having a total mass of $2.7 M_\odot$ -- as dominate the observed BNS population in the Milky Way (see~\cite{fBNS-prl25} for discussion and references) -- find time delays up to 10 s depending on the equation of state; additionally, the 1.7$\,$s between the observations of the gravitational wave GW170817 and the associated GRB may be emblematic of this time delay.    
The jets have 100 times larger Poynting flux and 6\% of the total solid angle, so if they contributed maximally to UHECR production they would contribute 6 times as much as the outflow. 
Therefore, we write the UHECR energy per merger as
\begin{equation}
\label{eq:Eest}
  E_{\rm equi}^{\rm UHECR} \approx 3.4 \,(f_{\rm ej} + 6 \, f_{\rm jet}) \!\left(\frac{T}{1.7{\rm \,s}
  }\right)\! \left(\frac{F_{\rm Poyn}}{10^{49.5}\rm{erg/sr/s}}\right) \! 10^{50} \rm{erg},    
\end{equation}
where $f_{\rm ej},f_{\rm jet} \leq 1$ are the fractions of energy in the magnetized turbulent ejecta and the jet going into UHECRs, respectively, relative to the maximum set by equipartition with the initial magnetic field energy.

We ask whether this estimate for the energy in UHECRs is consistent with that required to explain the overall normalization of the observed spectrum, given constraints on the BNS merger and UHECR energy injection rates.  The total energy of UHECRs accelerated in an individual merger, if BNS mergers account for all UHECRs, can be deduced from the volumetric energy injection rate in UHECRs ($\dot{\mathcal{Q}} \equiv \dot{\mathcal{Q}}_{45}\, 10^{45}\,\rm{erg \, Mpc^{-3} \, yr^{-1}} $) and the BNS merger rate, which we write as $\Gamma_{\rm BNS} = \Gamma_{3}\, 10^{3}\, \rm{Gpc^{-3} \, yr^{-1}} $.  Then the UHECR energy per merger required for BNS mergers to be the sole source of UHECRs is
\begin{equation}
\label{eq:Ereq}
    E_{\rm req}^{\rm UHECR} = \dot{\mathcal{Q}}/\Gamma_{\rm BNS} = (\dot{\mathcal{Q}}_{45} / \Gamma_3) \,10^{51} \,\rm{erg}.
\end{equation}
The LIGO-Virgo-Kagra collaboration's conservative range on the BNS merger rate is 10-1700 $\rm{Gpc^{-3} \, yr^{-1}} $~\citep{LVK_binaryMergerRate23}, while the recent detailed study of the afterglows of 29 short GRBs of~\citet{SGRBrate23} finds, using events with well-measured opening angles, a beaming-corrected short GRB rate of 
$1786^{+6346}_{-1507}\,\rm{Gpc^{-3} \, yr^{-1}}$. Combining these and taking the $1\sigma$ lower value from the latter, gives the estimate $0.3 \lesssim \Gamma_3 \lesssim 1.7$.  

Determining $\dot{\mathcal{Q}}$ from UHECR data is non-trivial because below 8 EeV there is a mixture of Galactic and extragalactic CRs and the extracted extragalactic (UHECR) component is sensitive to the assumed spectral shape at low energy.  A typical estimate is $ \dot{\mathcal{Q}}_{45}\approx \mathcal{O}(1) $; a more precise estimate of $\dot{\mathcal{Q}}_{45}$ requires fitting UHECR data using the spectral shape of escaping UHECRs appropriate for the BNS merger scenario (see Sec.~\ref{sec:escape}).  When the suppressed escape probability at low energies is properly taken into account in fitting the observations, the value of $ \dot{\mathcal{Q}}_{45}$ may prove to be considerably lower.  Even $\dot{\mathcal{Q}}_{45}=1$ is compatible with Eqs.~\eqref{eq:Eest} and~\eqref{eq:Ereq} within present uncertainties, the requirement being $\dot{\mathcal{Q}}_{45} = \frac13 L_{49.5}\, \Gamma_3 \, T_{1.7} (f_{\rm ej} + 6 \, f_{\rm jet})$.

\vspace{0.05in}
\section{Signatures of UHECRs' BNS-merger origin}
\label{sec:probe}

\emph{Neutrinos:}\\ A UHECR produced in a given BNS merger is deflected by the intervening magnetic fields and arrives with a characteristic time delay of $
    \tau \approx 0.14 \, (D_{\textrm Mpc} \beta_{\text{EGMF}}/\mathcal{R}_{\rm EV})^2 \,{\textrm{Myr}},$
where $\beta_\mathrm{EGMF} \equiv B_\mathrm{EGMF} / \mathrm{nG}\,\sqrt{L_c / \mathrm{Mpc}}
$ is the extragalactic magnetic smearing parameter~\citep{Achterberg99}.  Typical estimates are $B_\mathrm{EGMF}\approx 1 $ nG, and $L_c$ in the range 0.01-1 Mpc, giving $\beta_\mathrm{EGMF}\approx 0.1 - 1$; see~\citet{bf23} for further discussion.  Thus UHECRs arrive long after the gravitational wave associated with their production, 
defeating multi-messenger coincidences with UHECRs as a useful technique.  

However UHECRs make neutrinos, whose time delay relative to the arrival of the gravitational wave is the sum of the time needed to accelerate the UHECR and for it to produce the neutrino.  
So-called cosmogenic neutrinos are those produced in UHECR interactions with the CMB and other extragalactic background light after the UHECR has left its source environment.  Even though the speed of the UHECR is essentially the same as that of the GW,  its deflections prior to production of a cosmogenic neutrino, which occurs anywhere along the trajectory, result in $>$ kyr time delays and up to several degrees difference in direction between a cosmogenic neutrino and the GW associated with the UHECR's production.  Thus cosmogenic neutrino correlations are not individually useful for identifying the source of UHECRs.

We therefore focus on neutrinos produced while the UHECR is still within the merger ejecta and its associated photon field.  Such neutrinos are produced in two ways:  beta-decay of UHE neutrons produced by spallation or photo-disintegration, and decay of pions produced by photoproduction when UHE nuclei interact with the radiation field.  The peak energy of beta-decay neutrinos is $0.8 \times 10^{-3}$ times the energy of the neutron, while the peak energy of neutrinos from photo-pion production is 1/20th of the UHECR's energy per nucleon.  These neutrinos travel in the same direction as the nucleon due to relativistic beaming and their abundance is determined by the temperature of the photon field and density of the gas traversed by the UHECR nuclei.  See~\citet{UFA15} and~\citet{mfu22} for more detailed discussion. 

The spectrum of neutrons in nuclei having equal numbers of protons and neutrons, extends up to $E_{\rm cut}/A = \mathcal{R}_{\rm cut}/2 \approx 3-3.5$ EeV, while the nucleons in Fe and Te have spectral cutoffs at $E_{\rm cut}/A = 3.2$ and 2.7 EeV, respectively.  Once produced, a free neutron escapes directly and produces a beta-decay $\bar{\nu}_e$ with energy up to a few PeV.  Both neutron and neutrino travel in a straight line at essentially the speed of light.   
Photo-pion produced neutrinos have energies about 60 times higher, i.e., up to a maximum of $\approx 175$ PeV. (Energetically speaking, the KM3Net event~\citep{km3net25} could be such a photo-pion produced neutrino, although the effective exposure of IceCube in the direction of the KM3Net event is more than an order of magnitude higher than that of KM3Net so in any scenario the KM3Net detection and IceCube non-detection must be a statistical fluke.)  If there is a higher energy p or He component produced in the jet, as discussed in Sec.~\ref{sec:jet}, with energy per nucleon up to 10 EeV or 7.5 EeV respectively, the neutrinos they produce would be correspondingly more energetic, i.e., by roughly a factor of three.

Predicting the number and detailed spectrum of the EHE neutrinos from UHECRs produced in a BNS merger will require numerical simulation.  The highest energy neutrinos will be produced most abundantly shortly after production of the highest energy nuclei, at $r\approx 10^{14}$ cm, before the interaction rate drops due to dilution of the radiation field with continued expansion.  When simulations of the photon field in the merger outflow are extended to $ 10^{14}$ cm and beyond, the neutrino spectrum can be quantitatively calculated and the coincidence rate of EHE neutrinos with gravitational waves and short GRBs can be predicted. 

The sensitivity of present neutrino, GW and short gamma ray burst observatories is too low for a significant  rate of detection of coincidences, but ~\citet{fBNS-prl25} estimated very roughly that the coincidence rate could be of order one per year with IceCube Gen-2 and the proposed Cosmic Explorer and Einstein GW telescopes.  Taking the initial outflow to have a velocity 0.1 or 0.2 $c$, the time delay between gravitational wave and neutrino arrivals is several hours to a day, but can be longer if the parent UHECR is trapped in the merger outflow for some time before production of the neutron or neutrino.  Since the time required for a UHECR to escape from the ejecta increases with decreasing rigidity, the delay time between GW and neutrino arrival will characteristically be larger for lower energy neutrinos. 

\emph{Composition I: CRs heavier than Fe}\\
The AugerPrime upgrade completed in 2023 enables significantly more detailed measurements of individual UHECR air showers than previously possible.  This should permit better hadronic interaction models to be developed and solve the long-standing ``muon problem''~\citep{augerJeffTopDown} and the more recently identified ``$X_{\rm max}$ problem''~\citep{augerJakubPRD24}. With this better understanding, the mapping from the measured air shower to primary composition will be both more accurate and more precise. It will become possible to assign a composition PDF for each event.  Establishing the existence of UHECRs having A>56 would be strong evidence for their origin in the outflow of a BNS merger, since the abundance of such heavy elements is miniscule in nearly all other environments.  Already, the existence of two events above 200 EeV and many above 150 EeV is a strong hint that such an $A>56$ population exists~\citep{fBNS-prl25}.

\emph{Composition II: (non)Peters cycle as a probe}\\
With adequate statistics and sufficiently accurate mass and energy determination, the locus of production of oxygen group and helium can be probed via their spectrum.  If the deduced \Rcut\ values are higher than for heavier elements (taking into account the impact of interactions in the extragalactic journey to Earth), we can infer that these elements were accelerated at $\approx 10^{14}$ cm and hence must have been produced from breakup of the heavy initial nuclei prior to acceleration to UHE, as outlined in Sec.~\ref{sec:composition}.  If, instead, O and He are found to have a similar or lower value of \Rcut\ than the heavier nuclei, it would imply they are mostly remnants of post-acceleration breakup of heavier nuclei.  Even if there is direct acceleration of O and He, breakup of heavier nuclei leading to CRs with lower \Rcut\ will also be present, making this a challenging analysis to execute and interpret.  When the photon field beyond $10^{14}\,$cm has been determined, as is in progress (K. Kiuchi, private communication), it will become possible to self-consistently calculate the post-acceleration production of intermediate mass UHECRs.  

Protons from outside the jet region are the product of breakup of heavier nuclei, so they characteristically have a factor-2 lower energy than if they had been directly accelerated.  The typical energy of spallation-produced nucleons decreases with the increasing mass of the parent nucleus:  a nucleon spalled from Te having rigidity $\mathcal{R}$, has energy $0.4 \mathcal{R}$, to be compared to $0.46 \mathcal{R}$ for nucleons from Fe and $0.5\mathcal{R}$ for lighter parents.  Spallation interactions -- the primary energy-loss mechanism of ordinary nuclei -- conserve the energy per nucleon of the UHECR, whereas Bethe-Heitler interactions ($\gamma A \rightarrow e^+ e^- A$) reduce the energy per nucleon. The Bethe-Heitler cross section increases strongly with $Z$, such that for r-process nuclei it dominates spallation in some energy regimes and requires more detailed treatment than when the energy per nucleon is preserved during extragalactic propagation. See~\citet{ZhangMuraseUH24} for a discussion.  

\section{Future Work}
\label{sec:disc}

In previous sections we outlined how to predict the spectrum and composition of ultrahigh energy cosmic rays in the BNS merger scenario, and made simple analytic estimates.  The next step is to carry out more detailed modeling.  Particularly high priority studies are the following: \\ 
$\bullet$ \emph{PIC simulations of acceleration in spherically expanding magnetized turbulence} are needed to determine the appropriate prefactors in Eqs.~\eqref{eq:Rcut} and~\eqref{eq:tauacc} which may differ from those used here based on the fixed rectilinear geometry studied by~\citet{cfm24}.\\
$\bullet$ \emph{Determine the spectrum of synchrotron radiation} from CRs prior to reaching $r_{\rm crit}$, by including synchrotron radiation in these PIC simulations.  This will enable synchrotron losses to be included in the energy budget calculation, and show whether there may be a detectable EM signature of the accelerating cosmic rays.  \\ 
$\bullet$ \emph{Simulations of the UHECRs' escape from the magnetized outflow} including their interactions with the ambient photon field, to determine both the observed spectrum and composition and the level of EHE neutrino production~\citep{UFA15,muf19,mfu22}.\\
$\bullet$ \emph{PIC simulations of the uptake of nuclei} into the accelerating chain, to determine how the uptake depends on $Z,A$, local density, temperature and the magnetic field strength and structure.   This is necessary to make realistic composition predictions and to predict the total mass in UHECRs per merger event. ~\citet{Comisso19}'s study for an electron-positron plasma suggests reconnection is essential here; analogous studies are needed for electron-ion plasmas, for both magnetized turbulence and in shocks of different kinds. \\
$\bullet$ \emph{Predicting the relative abundances of intermediate and heavy UHECR nuclei:}  The initial composition distribution of nuclei will become well-determined, once the duration of the hypermassive neutron star prior to its collapse to a black hole is better known and other current uncertainties on nucleosynthesis are reduced.  In combination with a robust understanding of uptake into the accelerator as the outflow expands, this will provide the necessary foundation for predicting the UHECR composition distribution.  Once an ion's kinetic energy exceeds $\gtrsim 100\,$MeV, collisions with bulk nuclei in the still-quite-dense medium can cause breakup so that the initial composition, consisting of $Z,N>20$ nuclei, evolves into a spectrum of nuclei with a range of intermediate as well as heavy masses.  Ions surviving this phase of the expansion 
 with minimal spallation 
 become the very highest energy cosmic rays, with initial energies up to $Z \,\mathcal{R}_{\rm cut}$, i.e., $\approx 300$ EeV for nuclei in the first r-process Xe-Te peak.    

\section{Summary}
\label{sec:summary}
We have presented here the first quantitative, parameter-free prediction of the spectral cutoff of the various nuclear mass components of ultrahigh energy cosmic rays, finding good agreement with observations.  
Because the magnetic field produced in a binary neutron star merger is generated by a gravitationally-driven dynamo~\citep{Kiuchi24}, and the mass range of known binary neutron star systems is very narrow ($2.7 \, M_\odot \pm 5\%$), BNS mergers naturally explain~\citep{fBNS-prl25} the ``standard source'' property observed in the data.\footnote{
Even a factor-two range of BNS total masses, which translates to a factor-1.4 range in magnetic field strength, is compatible with the observed rigidity distribution of UHECRs.
 }
By contrast, source candidates such as AGN, long GRBs and TDEs, having a broad distribution of luminosities over the source population, do not~\citep{foteini+23}. 

The observed presence of a significant component of intermediate and heavy nuclei at the highest energies~\citep{augerSDXmax17,augerDNNPRL25,augerDNNPRD25}, combined with the results of nucleosynthesis studies showing that elements heavier than helium are only produced at larger angles outside the jets, rules out UHECR production being exclusively in the jets or their sheaths. (For a contrary view see ~\citet{guo+25}.)  Using the initial conditions on the magnetic field from the neutrino-GRMHD simulation of~\citet{Kiuchi24}, we showed that the outflow at larger polar angles outside the jets satisfies the criterion for  magnetized turbulence.  Adding the constraint of homologous expansion, established by~\citet{Rosswog+BNS14}, determines the evolution of the outflow and enables us to calculate the cutoff rigidity for particles accelerated therein.  

The spectrum produced by magnetized turbulence has been determined in PIC simulations to have the form $\mathcal{R}^{-s}\, {\rm sech}[(\mathcal{R}/\mathcal{R}_{\rm cut})^2]$~\citep{cfm24}.  
 Our no-free-parameter prediction, \Rcut$=6-7$ EV, except for helium, is in good agreement with the value determined by fitting UHECR data:  
$\mathcal{R}_{\rm cut} = 6.0\pm 0.1 \rm(stat.)\pm 0.8 \rm(syst.)\,$EV~\citep{augerCombFitJCAP23} and $\mathcal{R}_{\rm cut} = 6.3^{+6.3}_{-2.3}\,$EV~\citep{cfm24}.
If helium is present when the outflow has expanded to the locus of maximal UHECR acceleration, $r \approx 10^{14}\,$cm, we predict $\mathcal{R}_{\rm cut,He}\approx9\,$EV.  But if helium is produced later by spallation from heavier nuclei, it will have their characteristically lower energy per nucleon and a maximum rigidity $\lesssim 6\,$EV. 

We conjectured that after the jet has expanded to a radius $r \approx 10^{15}\,$cm, its magnetic field can be approximated by magnetized turbulence.  In that case the cutoff energies for p and He in the jet region can be calculated as for the large angle outflow, giving $E_{\rm cut} = 11.5$ and 35 EV, respectively.  (The abundance of p and He in the jet is uncertain because it is sensitive to the neutron star equation of state~\citep{Perego+LightElems22}.) Interestingly, the phenomenological study of UHECR data by~\citet{mfu22} found that inclusion of a light component at distinctly higher energy per nucleon than in the bulk of the CRs, carrying about 10\% of the energy above 10 EeV, improved the fit at $\approx 5\, \sigma$ confidence level. Such a light component may naturally be explained as due to p and/or He acceleration in the jet or its sheath.  

With future better composition information expected soon from the upgraded Auger Phase II observations and improved hadronic interaction models, multiple contributions to UHE protons and helium can potentially be separated, and thus help determine the neutron star equation of state and the locus of production of the various UHECR nuclei.  Our analysis of the BNS merger scenario calls for refitting the latest UHECR data using the sech cutoff and allowing for the more subtle pattern of rigidity cutoffs reported here. 

Although important work remains to develop more accurate predictions, the good consistency between the simple, parameter-free prediction of the spectral cutoff derived here and the value from UHECR observations adds to the evidence that BNS mergers can be the source of ultrahigh energy cosmic rays.  Definitive tests will be detection of neutrinos with PeV energies and above, with a correlated gravitational wave or short GRB occurring about a day earlier, and unambiguous demonstration that the highest energy cosmic rays have mass greater than iron.  

\section*{Acknowledgments}
I am grateful to K. Kiuchi, A. MacFadyen, E. Most, D. Radice, A. Grosberg, B. Metzger, L. Comisso, A. Ferrari, E. Guttierez, K. Murase, M. Muzio, F. Oikonomou, A. Spitkovsky, M. Unger
 and colleagues in the Pierre Auger Collaboration, for valuable input which has stimulated and enriched the work presented here. I also thank the anonymous referee whose helpful comments have improved the presentation. 
This research has been supported by National Science Foundation grant PHY-2413153 and the paper was completed at the Aspen Center for Physics, which is supported by National Science Foundation grant PHY-2210452.  

\appendix
\section{Synchrotron cooling \& Cutoff Rigidity}
\label{app:synch}
Acceleration of relativistic particles in magnetized turbulence is maximally efficient, with an acceleration timescale 
$(\dot{E}/E)^{-1}$ found by~\citet{cfm24} to be given by
\begin{equation}
\label{eq:tauacc}
t_{\rm acc} \approx 1.6 \, l_{\rm coh} /c~.
\end{equation} 
However in our application, the initial field strength is large and ion energies grow much more slowly than given by~\eqref{eq:tauacc} because the growth of energies is limited by synchrotron radiation.  The power in  synchrotron radiation is proportional to the square of the energy times the field strength:
\begin{equation}
\label{eq:synch}
    \dot{E}_{\rm synch} = \frac49 c \left(\frac{m_e r_0 }{ m_p}\right)^2 (\gamma \beta B)^2 \,\frac{Z^4}{A^2} ~,
\end{equation}
where $r_0$ is the classical radius of the electron, $\beta=v/c$, $\gamma = 1/\sqrt{(1 - \beta^2} = E/(A m_p)$ for relativistic ions, and $Z$ and $A$ are the charge and atomic mass number of the ion.   
As the ejecta expands, $\dot{E}_{\rm synch}$ for a given species and energy drops as $B^2\sim r^{-3}$.  The critical field strength below which synchrotron losses are subdominant at a given rigidity depends on $\{Z,A\}$ of the particle.  

\begin{figure}[!]
 \includegraphics[trim={0 0 0 0},clip,width=0.48\textwidth]{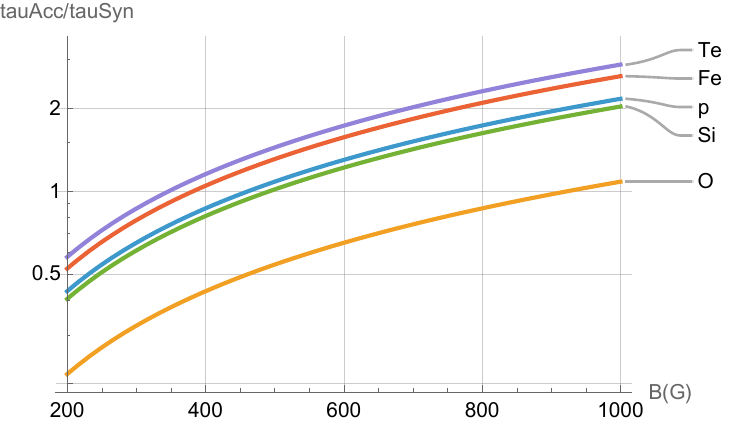}
    \caption{Acceleration timescale divided by synchrotron loss timescale versus rms magnetic field strength, for various nuclei, at \Rcut=6.8 EV; He is a factor 4 below O and crosses unity at 3700G.  \label{fig:SynchoverAcc} }
\end{figure}

When acceleration is not inhibited by synchrotron emission, the field strength at which a specified \Rcut\ is reached for a given $l_{\rm coh}$ is given by Eq.~\eqref{eq:Rcut}.  At the same time, $l_{\rm coh}$, determined in Sec.~\ref{sec:MagTurbAcc} to be $\approx r/3$, determines the acceleration timescale $t_{\rm acc}$ via Eq.~\eqref{eq:tauacc}.  Figure~\ref{fig:SynchoverAcc} shows the ratio $t_{\rm acc}/t_{\rm syn}$ as a function of $B$ at the central value of the cutoff rigidity derived from Auger data, \Rcut = 6.8 EV~\citep{augerCombFitJCAP23,cfm24}, for several nuclear species, with $t_{\rm syn} \equiv E/\dot{E}_{\rm synch}$.  In order for acceleration not to be synchrotron limited requires $t_{\rm acc}/t_{\rm syn} \lesssim 1$.

\vspace{0.05in}
\section{Escape from the source}
\label{app:escape}

As a crude first estimate, we can approximate the spectrum of escaping CRs of a given $Z,A$ by integrating the spectrum applicable inside the magnetized turbulence when its radius is $r$, $dN/d\,{\rm log}_{10}\, \mathcal{R}(r)$, weighted by the mean escape rate at radius $r$, $t_{\rm esc}(r,\mathcal{R})^{-1}$:  
\begin{equation}
\label{{fig:escSpec}}
    dN^{\rm esc}/d\,{\rm log}_{10} \mathcal{R} = v_{\rm ej}^{-1}\int dr\, dN/d\,{\rm log}_{10}\, \mathcal{R}(r)\, t_{\rm esc}^{-1}(r,\mathcal{R})~.
\end{equation}
Here $dN/d\, \mathcal{R}(r)$ is fixed by Eq.~\eqref{eq:specR} with $\mathcal{R}_{\rm cut}(r)$ given by the colored solid or dashed black line in Fig.~\ref{fig:synchlim} depending on whether $r$ is below or above $r_{\rm crit}$.

The escape time, $t_{\rm esc}(r,\mathcal{R})$, is the effective mean time for a CR of rigidity $\mathcal{R}$ to reach the surface of the ejecta when its radius is $r$, from a random initial starting position inside.  However this simple description hides a lot of complexity.  During the time it takes for a UHECR to escape, the diffusion coefficient is evolving because the coherence length is increasing and field strength dropping. The CRs whose Larmor radius is larger than the new coherence length, mainly escape with little energy transfer, while those at lower energy will diffuse in energy space~\citep{cfm24}.

\begin{figure}[t]
\includegraphics[width=0.48\textwidth, trim=0cm 0cm 0cm 0cm, clip=true]{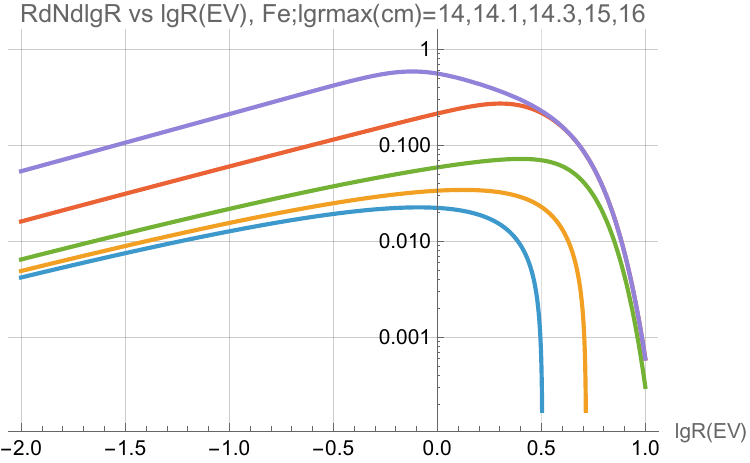}
    \caption{Rigidity-weighted escaping spectrum for Fe, $\mathcal{R}\, dN^{\rm esc}/d\,{\rm log}_{10}\mathcal{R}$ (a.u.),  after expansion to progressively larger radii $10^{14},\,10^{14.1},10^{14.3},\,10^{15},\,10^{16}\,$cm (shown in blue, orange, green, red and purple), in the simple approximation outlined in App.~\ref{app:escape}; a more complete treatment leads to greater suppression of late-time and low energy contributions. }
    \label{fig:escSpec}
\end{figure}

    A complete treatment of these effects requires numerical simulations beyond the scope of the present paper, but we can obtain a qualitative idea of the spectrum of escaping UHECRs as follows.  As a first analytic approximation, we ignore the expansion and other effects and approximate $t_{\rm esc}(r,\mathcal{R})$ as the time it takes a CR of rigidity $\mathcal{R}$ randomly located inside a sphere of radius $r$ with diffusion coefficient $D(r,\mathcal{R})$, to escape.  The mean time of first passage through the surface of the sphere is given by $t_{\rm esc}=\frac{r^2}{15 D(r,\mathcal{R})}$.  (See, e.g.,~\citet{RednerFirstPassageReview22} for a discussion of the theory of first passage and relationships useful for deriving the numerical coefficient, 15.) 

The diffusion coefficient for a UHECR of rigidity $\mathcal{R}$ in a Kolmogorov random field with rms strength $B$ and coherence length $l_{\rm coh}$ is given in~\citet{mf23} based on fitting the simulations of~\cite{Globus+07}. Evolving the~\citet{Kiuchi24} field initialization to the UHECR acceleration region determines the diffusion coefficient and escape time, revealing that typically $t_{\rm esc} \sim \mathcal{R}^{-1}$, hardening the spectrum by roughly one unit.
Figure~\ref{fig:escSpec} shows the resultant rigidity per dex of escaped CRs, $\mathcal{R}\, dN^{\rm esc}/d\,{\rm log}_{10}\mathcal{R}$, for Fe after different stages of expansion, assuming $B\sim r^{-3/2}$. Near $r_{\rm crit}$, where \Rcut\ is maximum for each species   ($10^{14.1}\,$cm for Fe), one sees formation of the characteristic spectral shape.  As $r$ increases past $r_{\rm crit}$, the high-rigidity end initially continues to be enhanced due to particles accelerated to \Rcut\ at smaller $r$ which take time to escape.  

At $r>> r_{\rm crit}$ (red and purple lines) the maximum rigidity of accelerated particles decreases as per the black dashed line in Fig.~\ref{fig:synchlim}. The contributions from these larger $r$'s are potentially further suppressed if, as seen by~\citet{cfm24} in their static-box simulations, a sufficiently large number of CRs are accelerated to high enough energy.  In that study, the total kinetic energy in CRs becomes comparable to the energy in the initial magnetic field and $\approx 50$\% of the initial magnetic energy was found to be transferred to the ions in steady state.  This behavior is consistent with naive equipartition, so it may be applicable here as well.  If so, feedback from energy transfer to CRs will cause the magnetic field strength to drop faster than $r^{-3/2}$, distorting the evolution of the spectrum relative to the simple picture discussed above.  The quantitative impact in practice depends on the efficiency of the uptake of ions into the acceleration chain because the total energy of UHECRs of any given species is the average UHECR energy for that species times the number of CRs of that species.  Understanding the uptake efficiency of CRs of a given species and injection into the regime of turbulent acceleration, which in general depends on $\{Z,A\}$, is therefore a high priority topic to study with PIC simulations.


\bibliographystyle{aasjournal}

\end{document}